\documentclass[USenglish]{article}
\usepackage[big,online]{dgruyter}
\usepackage{xcolor}

\begin{document}


  \author[1]{Dani Rodríguez-Castellanos}
  \author*[2]{Petr Jizba}
  \runningauthor{D. Rodríguez-Castellanos and P. Jizba}
  \affil[1]{FNSPE, Czech Technical University in Prague,
B\v{r}ehov\'a 7, 115 19 Prague, Czech Republic, E-mail: castedan@cvut.cz,
https://orcid.org/0000-0003-1980-4503}
  \affil[2]{FNSPE, Czech Technical University in Prague,
B\v{r}ehov\'a 7, 115 19 Prague, Czech Republic, E-mail: p.jizba@fjfi.cvut.cz,
https://orcid.org/0000-0002-7940-204X}
  \title{Thermodynamic Framework for $q$-Affinity}
  \runningtitle{Thermodynamic Framework for $q$-Affinity}
  %
\abstract{
We develop a thermodynamic framework for non-equilibrium affinities based on generalized entropies. In particular, we extend the classical concept of De Donder by introducing $q$-affinities associated with Rényi and Tsallis entropies. This in turn allows us to generalize thermodynamic driving forces to systems with long-range interactions and/or strong correlations. For Rényi entropy, we build on a thermodynamic interpretation due to Baez, where the entropy is expressed through finite differences of the Helmholtz free energy at two temperatures. This leads to a generalized thermodynamic potential whose derivative with respect to a reaction coordinate defines the Rényi $q$-affinity. The resulting expression admits a representation in terms of exponential work averages, establishing a connection to Jarzynski-type fluctuation relations. For Tsallis entropy, we consider Markov jump processes using a master-equation-based approach. We derive a $q$-deformed entropy balance law and obtain an explicit expression for the Tsallis entropy production rate, proving its non-negativity and thus recovering a generalized second-law structure. This allows to identify a local stochastic $q$-affinity with the generalized thermodynamic force entering the entropy production rate. 

  }
  \received{...}
  \journalname{J. Non-Equilib. Thermodyn. }
  \journalyear{2026}
 \DOI{...}

\maketitle

\section{Introduction}

In classical thermodynamics, the concept of \emph{affinity}, introduced by De Donder and Van Rysselberghe~\cite{dedonder1936thermodynamic}, quantifies the driving force of chemical reactions and plays a central role in non-equilibrium thermodynamics~\cite{pekar2020thermodynamic}. It provides a quantitative measure of the thermodynamic driving force underlying chemical reactions, characterizes the tendency of a system to evolve toward equilibrium and directly relates reaction progress to entropy production~\cite{prigogine1967irreversible}. Within the standard Gibbs--Shannon framework, the affinity is defined as
\begin{equation}
A \ = \  -\left(\frac{\partial G}{\partial \xi}\right)_{T,p}
  \ = \  -\sum_i \nu_i \mu_i\, ,
\label{eq:classical_affinity}
\end{equation}
where $\xi$ denotes the extent of reaction, $\nu_i$ are the stoichiometric coefficients, and $\mu_i$ are the corresponding chemical potentials.
The associated entropy production satisfies
\begin{equation}
dS_i \ = \  \frac{A}{T}\, d\xi \ \ge \  0\, ,
\label{eq:entropy_production}
\end{equation}
which determines the direction of spontaneous evolution in accordance with the second law of thermodynamics.

However, many real-world systems of current interest --- including complex materials, active matter, biological systems, and chemical reaction networks --- exhibit statistical behavior that is not adequately captured within the standard Gibbsian framework.  In such systems, the underlying dynamics often involves long-range interactions, persistent memory effects, or operates far from equilibrium. As a consequence, one observes a variety of nonstandard phenomena, including anomalous transport, non-Gaussian and heavy-tailed probability distributions, anomalous large-deviation behavior, as well as multiscaling and intermittency~\cite{Jou,Lebon,Colemen,tsallis2009book,Thurner,naudts2011generalised}.
These features fall outside the domain of validity of conventional (extensive and/or additive) thermodynamics and call for a generalized statistical description. In this context, generalized entropy measures provide a natural and versatile framework for capturing non-Gibbsian statistics and the effects of strong correlations~\cite{tsallis2009book,Thurner,naudts2011generalised}. In particular, Rényi and Tsallis entropies have emerged as prominent and tractable extensions of the standard Gibbs--Shannon formalism, offering a flexible setting for the analysis of complex non-equilibrium systems~\cite{tsallis2009book,Thurner,naudts2011generalised,shannon1948mathematical,tsallis1988possible,beck2009superstatistics,jizba2004renyi}.

At the same time, stochastic thermodynamics identifies \emph{entropy production} as a central measure of irreversibility in non-equilibrium systems~\cite{seifert2012stochastic}. This framework provides a consistent description at the level of individual trajectories and reveals deep connections between thermodynamics, information processing, and feedback-controlled dynamics~\cite{parrondo2015information,Seifert:book}.  Within this setting, it
becomes essential to reformulate the thermodynamic forces conjugate to reaction coordinates in a manner that
is consistent with both the underlying non-equilibrium dynamics and the use of generalized entropy measures.
Such a reformulation is necessary to provide a coherent description of driving forces in systems that deviate
from the standard Gibbsian paradigm

Rényi and Tsallis entropies represent two closely related, yet thermodynamically distinct, generalizations of the Gibbs--Shannon entropy. Both are parameterized by a deformation parameter $q \geq 0$ and recover the standard Gibbs--Shannon form in the limit $q \to 1$. However, they give rise to fundamentally different thermodynamic structures due to their distinct normalization schemes and definitions of conjugate thermodynamic  variables~\cite{renyi1961,Abe}.  In particular, Rényi entropy (RE) is additive and, for systems with exponentially growing state spaces, extensive; it also naturally incorporates escort distributions, which play a central role in non-equilibrium statistical mechanics and information geometry~\cite{beck2009superstatistics,jizba2004renyi,amari2016information}.  By contrast, Tsallis entropy (TE) is intrinsically non-additive and, under the same exponentially state-spaces scaling, non-extensive~\cite{tsallis2009book,tsallis1988possible}.
These features imply that any consistent generalization of thermodynamic forces based on R\'{e}nyi or Tsallis entropies must account not only for the entropy deformation itself --- which encodes correlations in the systems~\cite{jizba19}, but also for the altered probabilistic structure induced by escort averaging. In other words, the generalized forces conjugate to reaction coordinates or other thermodynamic variables should be defined in a manner that consistently reflects both the modified entropy measure and the corresponding statistical weighting of microscopic states.

In this work, we develop a unified thermodynamic framework for generalized affinity based on both R\'enyi and Tsallis entropies. The presented approach extends the classical concept of De~Donder by identifying the thermodynamic force conjugate to the extent of reaction within a generalized-entropic setting, thus providing a consistent generalization of affinity beyond the Gibbsian paradigm. Within the Rényi formulation, we show that the ensuing $q$-affinity can be expressed as an escort-weighted logarithmic response of the underlying probability distribution. This form provides an explicit information-theoretic characterization of the thermodynamic driving force and highlights the role of deformed statistical weighting in non-equilibrium processes. Moreover, this structure establishes a natural connection to geometric formulations of non-equilibrium thermodynamics, in particular to generalized divergences and thermodynamic length concepts~\cite{crooks2007thermodynamiclength}.

We further provide a direct thermodynamic interpretation of the Rényi $q$-affinity by exploiting the representation of Rényi entropy as a finite difference of Helmholtz free energies evaluated at two distinct temperatures~\cite{baez2022renyi}. This construction gives rise to a generalized free-energy functional whose gradient with respect to the reaction coordinate defines the $q$-affinity. In this way, the thermodynamic driving force is naturally extended beyond the conventional infinitesimal (linear-response) regime to a finite-temperature deformation of the underlying free-energy landscape.
This perspective highlights that the Rényi $q$-affinity encodes a non-local response in temperature space, thereby providing a thermodynamically consistent description of systems operating beyond the standard Gibbsian framework.

In contrast, within the Tsallis formulation, we derive the corresponding $q$-affinity directly from the gradient of the entropy functional and show that it is related to the Rényi expression through a normalization factor given by $\sum_i p_i^q$. This relation elucidates the complementary roles of the two entropic formalisms and embeds them within a unified thermodynamic framework. To incorporate dynamical aspects, we formulate a $q$-deformed entropy balance for continuous-time Markov jump processes and derive an explicit expression for the associated entropy production rate. The resulting bilinear current-force form extends the standard framework of stochastic thermodynamics~\cite{seifert2012stochastic,esposito2010threefaces} and ensures the non-negativity of entropy production, thus preserving the fundamental structure of the second law

Overall, the proposed framework generalizes the classical notion of thermodynamic affinity to systems governed by generalized entropy measures, while preserving its fundamental role as the driving thermodynamic force behind irreversible processes.
In doing so, it provides a unified perspective that explicitly highlights how escort distributions, finite-temperature deformations of free-energy differences, and $q$-logarithmic forces collectively shape the non-equilibrium behavior of non-equilibrium systems. 


\section{Extending the Concept of Affinity}

To extend the classical notion of thermodynamic affinity to non-equilibrium systems --- particularly those that may exhibit non-additive and/or non-extensive behavior  --- we consider generalized entropy measures that go beyond the standard Gibbs--Shannon paradigm.  Two closely related and widely studied candidates (that will serve us here as testbed entropies) are the Rényi and Tsallis entropic functionals, which are defined respectively as~\cite{tsallis1988possible,renyi1961} (Boltzmann constant $k_B =1$)
\begin{equation}
S_q^{(R)} \ = \ \frac{1}{1-q}\ln\!\left( \sum_i p_i^{\,q} \right), 
\qquad
S_q^{(T)} \ = \  \frac{1}{q-1}\left( 1 \ - \  \sum_i p_i^{\,q} \right),
\label{2.3.kl}
\end{equation}
where $\{p_i\}$ denotes the probability distribution over microstates and $q \ge 0$ is the non-extensivity (or correlation) parameter.  Although these entropic functionals are monotonic functions of each other, they induce  fundamentally distinct thermodynamic structures when used to define conjugate forces and entropy production rates. In the following, we derive the corresponding $q$-affinities by differentiating each entropy with respect to the extent of reaction, highlighting both their shared conceptual origin and the structural differences that arise from their respective definitions.

\subsection{$q$-Affinity based on R\'enyi Entropy}

In the following, we denote by $A_q^{(R)}$ the $q$-affinity associated with RE and define it in direct analogy with the classical affinity introduced by De Donder and Van Rysselberghe~\cite{dedonder1936thermodynamic}, but explicitly incorporating the dependence of thermodynamic potentials on the non-extensivity parameter $q$. For systems described by RE, we define the corresponding entropy production rate as
\begin{equation}
\frac{d S_q^{(R)}}{dt} \ \equiv\ \dot S_q^{(R)}
\ = \ 
\frac{A_q^{(R)}}{T}\,\frac{d\xi}{dt}\, ,
\label{eq:renyi_entropy_production}
\end{equation}
where $\xi$ represents the extent of reaction (also known as degree of advancement)~\cite{callen1985thermodynamics}. By differentiating the RE with respect to $\xi$, we obtain
\begin{equation}
\frac{d S_q^{(R)}}{d\xi}
\ = \ 
\frac{q}{1-q}
\frac{\sum_i p_i^{\,q-1}\, \dfrac{dp_i}{d\xi}}
{\sum_i p_i^{\,q}}\, .
\label{eq:renyi_derivative}
\end{equation}
By comparing  Eq.~\eqref{eq:renyi_derivative}  with~(\ref{eq:renyi_entropy_production}), we see that  the RE-based $q$-affinity can be explicitly written in the form
\begin{equation}
A_q^{(R)}
\ = \ 
\frac{qT}{1-q}
\frac{\sum_i p_i^{\,q-1}\, \dfrac{dp_i}{d\xi}}
{\sum_i p_i^{\,q}}\, .
\label{eq:q_affinity_definition}
\end{equation}
Equivalently, this expression can also be rewritten 
as 
\begin{equation}
A_q^{(R)}
\ = \ 
\frac{qT}{1-q}\,
\mathbb{E}_q\!\left[
\frac{d}{d\xi}\ln p
\right]\, ,
\label{eq:q_affinity_escort}
\end{equation}
where $\mathbb{E}_q[\cdots]$ denotes averaging with respect to the escort distribution~\cite{beck1993thermodynamics}:
$
\rho_i^{(q)}  =   {p_i^{\,q}}/({\sum_j p_j^{\,q}})$.
Formally, the derivative of the log-likelihood function $p_i(\xi)$ with respect to the parameter $\xi$ is known as the \emph{score function} (or \emph{score})~\cite{Fisher:35}. Accordingly, $A_q^{(R)}$ can be interpreted as an escort expectation of the score function. Since, the score quantifies the local steepness of the log-likelihood (thus captures the sensitivity of the model to infinitesimal changes in $\xi$), it plays a central role in estimation theory, where it serves as an important diagnostic and inferential tool~\cite{Friedan}. In  addition, it is also often used in information geometry where it determines the Riemannian metric of the manifold through the Fisher information~\cite{amari2016information,Friedan}. For this reason, one can interpret $A_q^{(R)}$  as a generalized statistical force that quantifies the system's response to infinitesimal variations along the reaction coordinate $\xi$ in probability space. 
Specifically in the framework of stochastic thermodynamics, such forces arise naturally as conjugate quantities to currents driving the system away from equilibrium (see also Sec.~\ref{Sec.2.4.2}). In this context, the score function can be viewed as generating infinitesimal displacements on the statistical manifold of probability distributions, while its expectation determines the corresponding thermodynamic driving force along the reaction coordinate. The $q$-generalization introduced here modifies this structure by replacing the standard linear averaging with escort averaging, thus inducing a deformation of the underlying information-geometric metric. As a result, $A_q^{(R)}$ encodes a non-extensive generalization of the thermodynamic force, in which the effective sensitivity of the system is reshaped by the relative weighting of fluctuations in probability space. 

\subsection{Thermodynamic Meaning of the R\'{e}nyi $q$-Affinity}
Let us now turn to applications in non-equilibrium thermodynamics (NET). In the context of RE, there exists an interesting connection between the correlation parameter $q$ and NEC. Following the approach introduced in Ref.~\cite{baez2022renyi}, one can identify the R\'enyi parameter $q$ with a ratio of two temperatures, namely
$q \equiv  {T_0}/{T}$,
where $T_0$ denotes the temperature of a reference equilibrium Gibbs state, and $T = T_0/q$ defines an auxiliary temperature.  Equivalently, in terms of inverse temperatures
$\beta  =   q\,\beta_0$.
Consider now a system with Hamiltonian $H$ whose (discrete) spectrum is given by
$\sigma(H) = \{E_i(\xi)\}_{i=0}^{\infty}$, where the energy levels depend on a reaction coordinate $\xi$. At a fixed temperature $T_0$, the system is described by the canonical Gibbs distribution,
\begin{equation}
p_i(\xi) \ = \ \frac{e^{-\beta_0 E_i(\xi)}}{Z(\beta_0,\xi)}\, ,
\qquad
Z(\beta_0,\xi) \ \equiv \ Z(T_0,\xi) \ = \ \sum_{i =0}^{\infty} e^{-\beta_0 E_i(\xi)}\, .
\label{2.2.8.bn}
\end{equation}
RE of order $q$ for the distribution~(\ref{2.2.8.bn}) can be written as
\begin{equation}
S_q^{(R)}(\xi)
\ = \
\frac{1}{1-q}
\ln\!\left( \sum_i p_i^q \right)
\ = \
\frac{1}{1-q}
\Big[
\ln Z(\beta,\xi)
\ - \
q \ln Z(\beta_0,\xi)
\Big]\, .
\label{eq:renyi_partition}
\end{equation}
Introducing the Helmholtz free energy
\begin{equation}
F(T,\xi) \  = \  -T \ln Z(T,\xi)\, ,
\end{equation}
Eq.~\eqref{eq:renyi_partition} can be recast as
\begin{equation}
S_q^{(R)}(\xi)
\ = \
-\frac{F(T,\xi) \ - \ F(T_0,\xi)}{T \ - \ T_0}\, ,
\label{eq:baez_secant}
\end{equation}
which admits a direct thermodynamic interpretation:  RE 
corresponds to the finite-temperature secant of the free energy. In particular, in the limit $q \to 1$ (i.e., $T \to T_0$), this expression reduces to the standard Gibbs--Clausius thermodynamic entropy
\begin{equation}
\lim_{q\to 1} S_q^{(R)}(\xi)
\ = \
-\left.\frac{\partial F}{\partial T}\right|_{T_0}
\ = \
S(\xi)\, .
\end{equation}
Motivated by this structure, we introduce the generalized free-energy potential
\begin{equation}
\Phi_q(\xi)
\ = \
\frac{T_0}{T_0 \ - \ T}
\big[ F(T,\xi) \ - \ F(T_0,\xi) \big]\, .
\label{eq:phi_q}
\end{equation}
This allows us to write the Rényi {$q$-affinity} as the thermodynamic force conjugate to the extent of reaction
%
\begin{equation}
A_q^{(R)}(\xi)
\ = \
-\frac{\partial \Phi_q(\xi)}{\partial\xi}
\ = \
T_0\,\frac{\partial S_q^{(R)}(\xi)}{\partial\xi}\, .
\label{eq:q_affinity_baez}
\end{equation}
In the limit $q \to 1$, the Rényi $q$-affinity smoothly transits into the conventional thermodynamic affinity, which demonstrates the consistency of the present construction with ordinary thermodynamics.

From the point of view of {\em stochastic thermodynamics}~\cite{Seifert:book}, the finite-temperature structure underlying Eq.~\eqref{eq:phi_q} acquires a deeper interpretation in terms of work fluctuations along non-equilibrium trajectories.  In fact, consider a system initially in thermal equilibrium described by state variables $\{T,\xi,X\}$ and equilibrium free energy $F(T,\xi,X)$. We perform a sudden quench by changing the heat bath (i.e. a thermal reservoir) temperature to $T_0$, thus driving the system out of equilibrium. Subsequently, we externally vary the control parameter (reaction coordinate) $\xi \to \xi'$ according to a prescribed protocol so that the corresponding equilibrium free energies satisfy
\begin{equation}
F(T_0,\xi',Y) \ = \  F(T,\xi, X)\, ,
\end{equation}
where both sides refer to equilibrium states of the system at different temperatures.

After the quench and driving protocol, the system is again driven in the heat bath by changing the reaction coordinate from $\xi'$ back to $\xi$, reaching a final state $\{T_0,\xi,Z\}$ with free energy $F(T_0,\xi,Z)$. In practice such a process is governed by entropy production and can be described within the framework of stochastic thermodynamics, e.g. via a non-equilibrium free energy functional. The associated free-energy difference 
\begin{eqnarray}
\Delta F \ = \ F(T, \xi, X) \ - \ F(T_0, \xi, Z) \ = \ F(T_0, \xi',Y) \ - \ F(T_0, \xi, Z) \, ,
\end{eqnarray}
can be related to the work performed on the system during the externally driven process $\{T_0, \xi',Y\} \to\{T_0, \xi, Z\}$ via the Jarzynski equality~\cite{Jarzynski:97,jarzynski1997nonequilibrium}, namely
\begin{eqnarray}
\Delta F \ = \ - T_0 \ln \left\langle e^{-\beta_0 W}\right\rangle  \;\;\; \Rightarrow \;\;\; \Phi_q(\xi)
\ = \ \frac{T_0^2}{T-T_0} \ \!\ln \left\langle e^{-\beta_0 W}\right\rangle\, ,
\end{eqnarray}
where $W$ is the work performed during the non-equilibrium realization of the protocol $\xi' \to \xi$, and $\langle\cdots\rangle$ denotes the ensemble average over many realizations starting from equilibrium at $\{T_0,\xi',Y\}$.

From this perspective, $A_q^{(R)}(\xi)$ encodes a temperature-space analogue of fluctuation relations, where the secant structure in Eq.~\eqref{eq:baez_secant} plays a role analogous to the cumulant-generating structure underlying Jarzynski-type identities. This places $A_q^{(R)}$ within a broader class of fluctuation-sensitive thermodynamic forces.
\subsection{$q$-Affinity based on Tsallis Entropy}

We now turn to Tsallis entropy and derive the corresponding generalized affinity using the same approach as for the R\'enyi-based $q$-affinity. The Tsallis entropy of order $q$ is defined by the second Eq.~(\ref{2.3.kl}), and it reduces to the Shannon entropy in the limit $q \to 1$.

By differentiating $S_q^{(T)}$ with respect to $\xi$, we obtain
\begin{equation}
\frac{d S_q^{(T)}(\xi)}{d\xi}
\ = \
-\frac{1}{q-1}\frac{d}{d\xi}\sum_i p_i^{\,q}
\ = \
\frac{q}{1-q}\sum_i p_i^{\,q-1}\frac{dp_i}{d\xi}\, .
\label{eq:tsallis_derivative}
\end{equation}
Analogous to the RE-based construction, we define the TE-based $q$-affinity by multiplying the derivative of $S_q^{(T)}$ with $T$, i.e.
\begin{equation}
A_q^{(T)}
\ \equiv \
T \frac{d S_q^{(T)}}{d\xi}
\ = \
\frac{T q}{1-q}
\sum_i p_i^{\,q-1}
\frac{dp_i}{d\xi}\, .
\label{eq:tsallis_q_affinity}
\end{equation}
In non-extensive statistical mechanics, the entropic index $q$ is not an arbitrary deformation parameter but reflects intrinsic properties of the underlying dynamics, such as long-range interactions, long-term memory, strong correlations, or non-ergodic exploration of phase space~\cite{tsallis2009book,Thurner,naudts2011generalised,tsallis1988possible}. 

It is worth noting that the RE-based $q$-affinity~(\ref{eq:q_affinity_escort}) is related to~\eqref{eq:tsallis_q_affinity} via the simple relation
\begin{equation}
A_q^{(R)}
\ = \
\frac{A_q^{(T)}}{\sum_i p_i^{\,q}}\, ,
\label{eq:renyi_tsallis_relation}
\end{equation}
which is a direct consequence of the fact that
\begin{eqnarray}
d S_q^{(T)} \ = \ \Big(\sum_i p_i^{\,q}\Big) \ \!d S_q^{(R)} \, ,
\end{eqnarray}
Eq.~\eqref{eq:renyi_tsallis_relation} shows that the R\'enyi $q$-affinity corresponds to a normalized version of the TE-based affinity, ensuring consistency with the information-geometric structure underlying non-extensive thermodynamics~\cite{jizba2004renyi,amari2016information}.

\subsection{Markovian Framework for Tsallis Entropy Production}

To build intuition for the TE-based $q$-affinity, we now analyze its behavior within a Markovian setting.

\subsubsection{Markov Jump Dynamics}

We consider a continuous-time Markov jump process (MJP) defined on a discrete set of states $m$~\cite{seifert2012stochastic,esposito2010threefaces}.
The probability distribution $p_m(t)$ evolves according to the master equation
\begin{equation}
\dot p_m(t)
\ = \
\sum_{m',\nu}
W^{(\nu)}_{m, m'}(\lambda_t)\, p_{m'}(t)\, ,
\label{eq:ME_24}
\end{equation}
where $W^{(\nu)}_{m, m'}(\lambda_t)$ denotes the transition rate from state
$m'$ to state $m$ associated with mechanism (reservoir) $\nu$, and
$\lambda_t$ is a possibly time-dependent external driving parameter. Note that~(\ref{eq:ME_24}) implies $\sum_{m, \nu} W^{(\nu)}_{m m'}(\lambda_t)  =  0$.

We define the probability current along the oriented edge $m' \to m$ as
\begin{equation}
J^{(\nu)}_{m, m'}(t)
\ \equiv \
W^{(\nu)}_{m ,m'}(\lambda_t)\, p_{m'}(t)
\ - \ 
W^{(\nu)}_{m', m}(\lambda_t)\, p_m(t)\, ,
\label{eq:current_24}
\end{equation}
which by definition obeys the antisymmetry relation
\begin{equation}
J^{(\nu)}_{m, m'} \ = \ -\, J^{(\nu)}_{m', m}\, .
\end{equation}
For convenience , we also introduce the forward and backward probability fluxes
\begin{equation}
a^{(\nu)}_{m m'}\ \equiv \  W^{(\nu)}_{m m'}\, p_{m'}\, ,
\qquad
b^{(\nu)}_{m m'} \ \equiv \  W^{(\nu)}_{m' m}\, p_m\, ,
\label{eq:fluxes_24}
\end{equation}
so that
\begin{equation}
J^{(\nu)}_{m ,m'} \ = \  a^{(\nu)}_{m, m'} \ - \  b^{(\nu)}_{m,m'}\, .
\label{23.kl}
\end{equation}
Let us now consider TE of order $q>0$ and compute its time derivative. By employing the master equation~\eqref{eq:ME_24}, together with the condition $\sum_{m, \nu} W^{(\nu)}_{m m'}(\lambda_t)  =  0$,  we obtain 
\begin{equation}
\dot{S}_q^{(T)}(t)
\ = \ 
\frac{q}{1-q}
\sum_m {p_m^{q-1}(t)}\ \! \dot p_m(t)
\ = \ 
\frac{1}{2}
\sum_{m,m',\nu}
J^{(\nu)}_{m, m'}(t)\,
Y_{q;m, m'}(t)\, ,
\label{eq:bilinear_24}
\end{equation}
where, we have defined
\begin{equation}
Y_{q;m, m'}(t)
\ = \ 
\frac{q}{1-q}
\left[
{p_m^{q-1}(t)}
\ - \ 
{p_{m'}^{q-1}(t)}
\right] \ = \ q\left[\ln_q \left(\frac{1}{p_m(t)}\right) \ - \   \ln_q \left(\frac{1}{p_{m'}(t)}\right) \right]\, . 
\label{eq:Yq_25}
\end{equation}
Here we have introduced the $q$-logarithm
\begin{equation}
\ln_q(x) \ = \ \frac{x^{1-q} \ - \ 1}{1-q}\, ,
\qquad x>0\, ,
\label{eq:qlog_24}
\end{equation}
which reduces to the standard logarithm in the limit  $q \to 1$. This definition also underlies the replica trick, widely used in the statistical physics of spin glasses and other systems with quenched disorder~\cite{parisi:95}.

For RE, an analogous expression can be derived, provided one replaces 
 $Y_{q;m m'} \mapsto \tilde{Y}_{q;m m'}$ where
\begin{equation}
\tilde{Y}_{q;m, m'}(t)
\ = \ 
\frac{q}{1-q}
\left[
\frac{p_m^{q-1}(t)}{\sum_k p_k^q(t)}
\ - \ 
\frac{p_{m'}^{q-1}(t)}{\sum_{k} p_k^q(t)}
\right] \, .
\label{eq:Yq_25}
\end{equation}

\subsubsection{$q$-Deformed Thermodynamic Force and Entropy Production Rate \label{Sec.2.4.2}}

We now construct the $q$-deformed entropy production rate in close analogy with conventional (Shannon-based) stochastic thermodynamics. 
To this end, we briefly recall the relevant relations from the conventional irreversible thermodynamics.

Consider, for instance, a network of chemical reactions treated as an open system, whose evolution generally proceeds out of equilibrium. The infinitesimal change in the system entropy can be decomposed as
\begin{equation}
dS_s \ = \ dS_{{e}} \ + \ dS_i \, ,
\label{3.3.aa}
\end{equation}
where $S_s$ denotes the entropy of the system, $dS_{{e}}$  represents the entropy exchange with the environment (e.g., a heat reservoir), and $dS_i$ is the entropy production due to irreversible processes within the system. 
In the particular case where the system exchanges heat $\delta Q$ with a reservoir at temperature $T$, the entropy flow is 
\begin{eqnarray}
 dS_{{e}} \ = \ \frac{\delta Q}{T}\, .
\end{eqnarray}
To extend the above thermodynamic framework to dynamical situations, one may introduce an explicit time dependence $t$, see e.g.~\cite{prigogine1955thermodynamics}.  In such a case, the entropy balance equation can be written in the rate form as
\begin{equation}
\frac{dS_s}{dt} \ = \  \frac{dS_{\small{e}} }{dt} \ + \  \frac{dS_i}{dt}\, ,
\label{3.35.cf}
\end{equation}
where the entropy-production rate $\dot{S}_i$ is always positive or zero. i.e. $\dot{S}_i  \geq  0$, 
%
%
with equality holding only for reversible processes. The latter is a direct consequence of the second law of thermodynamics.
Now, using Eqs.~(\ref{eq:bilinear_24}) and~(\ref{eq:Yq_25}), together with the identity
\begin{eqnarray}
\ln_q\left(\frac{x}{y}\right) \ - \ \ln_q\left(\frac{z}{v}\right) 
&=& \left[\ln_q\left(\frac{x}{z}\right)\right]\left(\frac{z}{v}\right)^{1-q} \ - \ \left[\ln_q\left(\frac{y}{v}\right)\right]\left(\frac{x}{y}\right)^{1-q}\, ,
\end{eqnarray}
and suppressing, for notational simplicity, the explicit time dependence of $p_m$ and $\lambda_t$ dependence of $W$, we can rewrite~(\ref{eq:bilinear_24}) in the form
\begin{eqnarray}
\dot{S}_q^{(T)} &=& 
\frac{1}{2}
\sum_{m,m',\nu}
J^{(\nu)}_{m, m'}\,
Y_{q;m, m'} \ = \ \frac{q}{2}
\sum_{m,m',\nu}
J^{(\nu)}_{m, m'} \left[\ln_q \left(\frac{1}{p_m}\right) \ - \   \ln_q \left(\frac{1}{p_{m'}}\right) \right] \nonumber \\[2mm]
&=& \frac{q}{2}
\sum_{m,m',\nu}
J^{(\nu)}_{m, m'} \left[\ln_q \left(\frac{W_{m',m}^{(\nu)}}{W_{m',m}^{(\nu)}\ \!p_m}\right) \ - \   \ln_q \left(\frac{W_{m,m'}^{(\nu)}}{W_{m,m'}^{(\nu)} \ \!p_{m'}}\right) \right] \nonumber \\[2mm]
&=& \frac{q}{2}
\sum_{m,m',\nu}
J^{(\nu)}_{m, m'} \left[p_{m'}^{q-1} \ \! \ln_q \left(\frac{W_{m',m}^{(\nu)}}{W_{m,m'}^{(\nu)}}\right) \ - \   p_{m}^{q-1} \ \! \ln_q \left(\frac{W_{m',m}^{(\nu)} \ \!p_{m}}{W_{m,m'}^{(\nu)} \ \!p_{m'}}\right)  \right] \, .
\label{eq:bilinear_30.kl}
\end{eqnarray}
With this representation, the entropy production rate can be written in a form closely analogous to that of standard irreversible thermodynamics, cf. Eq.~(\ref{3.35.cf}). In particular, the quantity
\begin{eqnarray}
\dot S^{(T)}_{e,q}(t) \ \equiv \
\frac{q}{2} \sum_{m,m',\nu}
J^{(\nu)}_{m, m'}(t) \ \! p_{m'}^{q-1}(t) \ \! \ln_q \left(\frac{W_{m',m}^{(\nu)}(\lambda_t)}{W_{m,m'}^{(\nu)}(\lambda_t)}\right) \, ,
\label{eq:34.ku}
\end{eqnarray}
can be identified with the TE flow. In the limit $q\rightarrow 1$, this expression correctly reduces to the corresponding result of conventional irreversible thermodynamics. On the other hand, the quantity
\begin{eqnarray}
\dot S^{(T)}_{i,q}(t) \ \equiv \
-\frac{q}{2} \sum_{m,m',\nu}
J^{(\nu)}_{m, m'}(t) \ \! p_{m}^{q-1}(t) \ \! \ln_q \left(\frac{W_{m',m}^{(\nu)} \ \!p_{m}}{W_{m,m'}^{(\nu)} \ \!p_{m'}}\right) \ = \  \frac{1}{2} \sum_{m,m',\nu}
J^{(\nu)}_{m, m'}(t) \ \! X^{(\nu)}_{q;m, m'}(t)\, ,
\label{eq:35.bb}
\end{eqnarray}
can be identified with the entropy-production rate. Here we have defined corresponding forces 
\begin{eqnarray}
X^{(\nu)}_{q;m, m'}(t)
\ = \ -q \ \! p_{m}^{q-1}(t) \ \! \ln_q \left(\frac{W_{m',m}^{(\nu)} \ \!p_{m}}{W_{m,m'}^{(\nu)} \ \!p_{m'}}\right)\, .
\end{eqnarray}
Apart from exhibiting the correct behavior in the limit  $q\rightarrow 1$, expression~(\ref{eq:35.bb}) is nonnegative for all admissible values of $q$. To demonstrate this, we rewrite~(\ref{eq:35.bb}) in the form
\begin{equation}
\dot S^{(T)}_{i,q}(t)
\ = \ 
-\frac{q}{2}
\sum_{m,m',\nu} p_m^{q-1}(t)
\big[a^{(\nu)}_{m m'}(t) \ - \ b^{(\nu)}_{m m'}(t)\big]
\ln_q\!\left(
\frac{b^{(\nu)}_{m m'}(t)}
     {a^{(\nu)}_{m m'}(t)}
\right)\, .
\label{eq:EP_explicit_36.kl}
\end{equation}
At this stage, we note that for $q>0$
\begin{equation}
\frac{d}{dx}\ln_q(x) \ = \ x^{-q} \ > \ 0\, ,
\label{eq:qlog_derivative_24}
\end{equation}
so the $q$-logarithm $\ln_q(x)$ is a strictly increasing function. Since $\ln_q(1) = 0$, we have that $\ln_q(a/b)$ (for any $a,b >0$) has the same sign as $a/b-1$, and thus
\begin{equation}
(a   - b)\ln_q\!\Big(\frac{a}{b}\Big) \ \ge \  0\, ,
\label{eq:positivity_term_24}
\end{equation}
with equality if and only if $a=b$. Consequently, we obtain
\begin{equation}
\dot S^{(T)}_{i,q}(t) \ \ge \ 0\, ,
\label{eq:entropy_positive_24}
\end{equation}
with equality holding if and only if
\begin{eqnarray}
W_{m',m}^{(\nu)} \ \!p_{m} \ = \ W_{m,m'}^{(\nu)} \ \!p_{m'}\, ,
\label{eq:41.kk}
\end{eqnarray}
i.e., when the condition of detailed balance is satisfied.
Thus, Eqs.~(\ref{eq:bilinear_30.kl}),~(\ref{eq:34.ku}), and~(\ref{eq:35.bb}) establish a consistent 
$q$-deformed generalization of the standard entropy balance structure. This balance can be expressed as
\begin{eqnarray}
\dot{S}_q^{(T)}(t) \ \equiv \ \dot{S}_{s,q}^{(T)}(t) \ = \ \dot{S}_{e,q}^{(T)}(t) \ + \ \dot{S}_{i,q}^{(T)}(t)\, .
\end{eqnarray}
Let us finally note that we can identify the generalized force $X^{(\nu)}_{q;m,m'}(t)$ with a local stochastic
$q$-affinity, i.e.
\begin{equation}
A^{(\nu)}_{q;m,m'}(t) \ \equiv \ X^{(\nu)}_{q;m,m'}(t).
\end{equation}
Accordingly, Eq.~(\ref{eq:35.bb}) can be rewritten in the compact current-affinity form
\begin{equation}
\dot S^{(T)}_{i,q}(t) \  = \ \frac{1}{2}\sum_{m,m',\nu}
J^{(\nu)}_{m,m'}(t)\,A^{(\nu)}_{q;m,m'}(t)\, .
\end{equation}
This representation shows that the TE framework preserves the universal force-flux structure of irreversible thermodynamics, with entropy production arising as the sum of local
probability currents multiplied by their conjugate generalized driving forces. In this sense, $A^{(\nu)}_{q;m,m'}$ measures the local departure from generalized detailed balance and
provides the natural extension of thermodynamic affinity to non-extensive stochastic systems.
Such a formulation is particularly relevant for complex networks in which irreversibility is distributed over many microscopic transitions, including chemical reaction networks, active matter, transport in disordered media, biological systems with memory, and anomalous diffusion processes. In these cases, the entropic index $q$ typically quantifies correlations in the system~\cite{jizba19,beck1993thermodynamics}. Although the above derivation was carried out for continuous-time MJP, many non-Markovian systems admit a Markovian embedding in an enlarged state space including auxiliary memory variables. In such cases, the local stochastic $q$-affinity remains well defined in the extended representation and induces an effective generalized affinity for the reduced non-Markovian dynamics.

\section{Conclusions}

By extending the classical De Donder concept, we have constructed Rényi- and Tsallis-entropy-based generalizations of thermodynamic affinity, thus generalizing the notion of thermodynamic driving forces to systems with complex statistical structure exhibiting long-range interactions or correlations. Examples of ensuing chemical systems include, e.g. ionic solutions with Coulomb interactions, hydrogen-bonded liquids, reaction--diffusion systems such as the Belousov--Zhabotinsky reaction or fluids near critical points.

Exploiting the finite-temperature representation of RE in terms of equilibrium free energies, we derived an explicit expression for the $q$-affinity, $A_q^{(R)}$, as the derivative of a generalized free-energy potential $\Phi_q$. Furthermore, we have shown that $\Phi_q$ admits a representation in terms of exponential work averages, which thus allowed to establish a direct link to fluctuation relations of Jarzynski type. This connection provides a non-equilibrium interpretation of $\Phi_q$ and identifies $A_q^{(R)}$ as a quantity sensitive to work fluctuations along driven processes. 

In the case of TE, we address the notion of $q$-affinity within the framework of MJP using a master-equation-based approach. In this setting, we derive a $q$-deformed entropy balance law and obtain an explicit expression for the TE production rate, establishing its non-negativity and thus confirming a generalized second-law structure. We further identify a local stochastic $q$-affinity associated with the corresponding generalized force entering the entropy production rate.

Results obtained demonstrate that both the Rényi and Tsallis frameworks provide consistent extensions of thermodynamic forces, with clear interpretations in both equilibrium and non-equilibrium settings. Moreover, all generalized expressions reduce smoothly to the classical thermodynamic results in the limit $q \to 1$, where Rényi and Tsallis entropies recover the Shannon entropy and the $q$-affinity converges to the standard thermodynamic affinity.


\begin{acknowledgement}
Both D.R.C. and P.J. were supported by the Czech Science Foundation Grant (GA\v{C}R), Grant No. 25-18105S. D.R.C. was also, in part, supported by the Grant Agency of the Czech Technical University in Prague, Grant No. SGS25/163/OHK4/3T/14.
\end{acknowledgement}


\end{document}